\documentclass[doublecol]{epl2}

\usepackage{amssymb}
\usepackage{amsmath,amssymb}
\usepackage{amsmath}

\title{Finite temperature and dissipative corrections to the Gross-Pitaevskii equation from $\lambda\Phi^4$ one loop contributions}
\author{T. Matos\inst{1} \and A. Su\'arez\inst{1}}
\institute{ \inst{1} Departamento de F\'isica, Centro de Investigaci\'on y de Estudios Avanzados del IPN, 07000 M\'exico D.F., M\'exico}

\abstract{Starting with a scalar field in a thermal bath and using the one loop quantum correction potential, we rewrite the Klein-Gordon equation in its thermodynamical representation and study the behavior of this scalar field due to temperature variations in the equations of motion. We find the generalization of a Gross-Pitaevskii like equation for a relativistic Bose gas with finite temperature, the corresponding thermodynamic and viscosity expressions, and an expression for the postulate of the first law of the thermodynamics for this BECs. We also propose that the equations obtained might help to explain at some level the phase transition of a Bose-Einstein Condensate in terms of quantum field theory in a simple way.}

\begin{document}
\maketitle

Phase transitions are changes of state, related with changes of symmetries of the system. The analisys of symmetry breaking (SB) mechanisms have turn out to be very helpful in the study of phenomena associated to phase transitions in almost all areas of physics. Bose-Einstein Condensation (BEC) is one topic of interest that uses in an extensive way the SB mechanisms \cite{b.c}, its phase transition associated with the condensation of atoms in the state of lowest energy and is the consequence of quantum, statistical and thermodynamical effects. The concept of BEC has in recent years emerged in an array of exciting new experimental and theoretical systems and is now a common phenomenon occurring in physics at all scales, from condensed matter to nuclear, elementary particle physics, astrophysics and now cosmology, with ideas flowing across boundaries between fields. The net results of these varied interests in this phenomenon has done that the emphasis gradually shifted from the study of nonrelativistic Bose systems to that of relativistic ones,\cite{b.v}. The increasing studies in the correspondence between SB in condensed matter physics and cosmology are one of the motivations of this research \cite{b.d,b.e,b.f}. Although phase transitions happen in condensed matter systems and the Universe, their status is not similar in both cases. For condensed matter, more transitions are known, for which our knowledge is better \cite{b.g,b.h,b.i}.\\

The results from finite temperature quantum field theory raise important challenges about their possible manifestation in condensed matter systems, which can be drawn by non-relativistic or relativistic scalar field (SF) theories in the framework of the Ginzburg-Landau theory of condensed systems. By investigating the massive Klein-Gordon equation we can be able, in principle, to simulate a condensed matter system. Since many particles have mass, this can be an essential step in building realistic analogue models. Here, our aim is to study a relativistic model made up of a real SF together with the possibility that this SF might undergo a phase transition as the temperature of the system is lowered. To analyze the SB undergone by the system we take a model having an U(1) symmetry, considering the easiest case of a double-well interacting potential (Mexican-hat potential) for a real SF $\Phi(\vec{x},t)$ that goes as
  \begin{equation}
   V(\Phi_c)=-\frac{1}{2}\frac{m^2c^2}{\hbar^2}\Phi^2+\frac{\lambda}{4\hbar^2c^2}\Phi^4.
  \label{eq:1}
  \end{equation}
Here one important idea, to which we shall refer, is that of identifying the order parameter $\Psi$ which characterizes the phase transition with the value of the real scalar quantum field $\Phi$.\\

From quantum field theory we know that the dynamics of a SF is governed by the Klein-Gordon (KG) equation, it is the equation of motion of a field composed of spinless particles. In this case we will add a first order auto-interaction potential to the SF such that the KG equation will be given by
 \begin{equation}
 \Box^2\Phi+\frac{\mathrm{d}V}{\mathrm{d}\Phi}-2\frac{m^2c^2}{\hbar^2}{\Phi}\phi=0,
 \label{eq:KG}
 \end{equation}
where the interaction potential will be denoted by $\phi$, and as will be seen later on it represents the gravitational potential, as usual the D'Alambertian operator is,
 \begin{equation}
  \Box^2\equiv-\frac{1}{c^2}\frac{\partial^2}{\partial t^2}+\nabla^2\label{eq:Box}\nonumber.
 \end{equation}\\
Here we will be interested in studying the properties, behavior and interactions of the SF with other particles, all of them interacting inside a thermal bath in a reservoir that can have interaction with its surroundings. In this case, the SF can be described by the potential (\ref{eq:1}) extended to one loop, see \cite{b.j,b.k,b.l}. In this case the finite-temperature effective potential denotes the free-energy density associated with the field $\Phi$, and goes as:
 \begin{equation}
  V_T(\Phi)=V(\Phi_c)+\frac{\hat\lambda}{8}k_B^2T^2\Phi^2 -\frac{\pi^2k_B^4}{90\hbar^2c^2}T^4,
 \label{eq:VT}
 \end{equation}
where $\hat\lambda=\lambda/(\hbar^2c^2)$ is the parameter describing the interaction, $\hat{m}^2=m^2c^2/\hbar^2$ is the scalar particle mass, $k_B$ is Boltzmann's constant, $\hbar$ is Planck's constant, $c$ is the speed of light and $T$ is the temperature of the thermal bath, this result includes both quantum and thermal contributions.\\

 From (\ref{eq:VT}), the energy density can be computed, $\rho_{\Phi}=V_T(\Phi)+Ts_{\Phi}$, where $s_{\Phi}=-\partial 
V_T(\Phi)/\partial T$. When the temperature $T$ is high enough, one of the minimums of the potential (\ref{eq:VT}) is $\Phi=0$. At this point, the SF density is equal to $\rho_\Phi\sim T_0T^4$, with $T_0=\pi^2k_B^4/30\hbar^2c^2$. We suppose that the temperature is sufficiently small so that the interaction between the SF and the rest of matter decouples. After this moment, the term $T_0T^4$ is not longer important, as for sufficiently low $T$ the term that goes as $T^4$ can be dropped out.\\

The critical temperature where the minimum of the potential $\Phi=0$ becomes a maximum and at which the symmetry is broken is, 
 \begin{equation}
  k_BT_c=\frac{2m}{\sqrt{\lambda}}.
 \label{eq:Tc}
 \end{equation}\\

Now for the SF we perform the transformation 
 \begin{equation}
  {\kappa}\Phi=\frac{1}{\sqrt{2}}\left(\Psi\,\mathrm{e}^{-\mathrm{i}\hat mct}+\Psi^*\,\mathrm{e}^{\mathrm{i}\hat mct}\right)\label{eq:Phi} 
  \nonumber,
 \end{equation}
where $\kappa$ is the scale of the system, which is to be determined by a experiment, from now on we will take $c=1$.\\

In terms of function $\Psi$, the KG equation (\ref{eq:KG}) reads,
\begin{eqnarray}
 \mathrm{i}\hbar\dot{\Psi}+\frac{\hbar^2}{2m}\Box^2{\Psi}+\frac{3\lambda}{2m}|\Psi|^2\Psi-m\phi\Psi
 +\frac{\lambda k_B^2T^2}{8m}\Psi=0,
\label{eq:Schrodinger}
\end{eqnarray}
where we have kept just the equation for the real part, the complex conjugate can be described in the same way. The notation used is; $\dot{}=\partial/\partial t$ and $|\Psi|^2=\Psi\Psi^*=\rho$. (\ref{eq:Schrodinger}) is KG's equation rewritten in terms of the function $\Psi$ and temperature $T$. This equation is an exact equation defining the field $\Psi(\mathbf{x},t)$, where $\phi$ defines the external potential acting on the system and the terms in $\lambda$ represent the interaction potential within the system.\\

 From the mathematical point of view, is a type of nonlinear Schr\"odinger equation. From now on equation (\ref{eq:Schrodinger}) will be 
considered as a generalization of the Gross-Pitaevskii equation for finite temperatures and relativistic particles. This is because at second order, when $T=0$ and in the non-relativistic limit, $\Box^2\rightarrow\nabla^2$, eq. (\ref{eq:Schrodinger}) becomes the Schr\"odinger equation with an extra term $|\Psi|^2$. In this limit this is known as the Gross-Pitaevskii equation for Bose-Einstein Condensates (BEC), i.e, an approximate equation for the mean-field order parameter in the classical theory, \cite{b.n}. The static limit of equation (\ref{eq:Schrodinger}) is known as the Ginzburg-Landau equation.\\

In what follows we transform the generalized Gross-Pitaevskii equation (\ref{eq:Schrodinger}) into its analogous hydrodynamical version, \cite{b.o,b.p}, for this purpose the ensemble wave function $\Psi$ will be represented in terms of a modulus $\hat\rho$ and a phase $S$ as,
 \begin{equation}
  \Psi=\sqrt{\hat{\rho}}\,\mathrm{e}^{\mathrm{i}S}.
 \label{eq:psi}
 \end{equation}
where the phase $S(\boldsymbol{x},t)$ is taken as a real function. As usual this phase will define the velocity. Here we will interpret $\hat\rho(\boldsymbol{x},t)=\rho/M_T$ as the rate between the number density of particles in the condensed state, $\rho=mn_0=mN_0/L^3$, being $N_0$ the number of particles in condensed state and $M_T $ the total mass of the particles in the system, being both, $S$ and $\hat{\rho}$, functions of time and position. The concept of SB is often used as a sufficient condition for BEC. The assumption that the ground state can be macroscopically occupied, is nothing but Einstein's criterion for condensation.\\

So, from this interpretation we have that when the KG equation oscillates around the $\Phi=0$ minimum, the number of particles in the ground state is zero, $\hat\rho=0=\rho$. Below the critical temperature $T_c$, close to the second minimum, $\Phi_{min}^2=k_B^2(T_c^2-T^2)/4$, the density will oscillate around $\hat\rho=\kappa^2k_B^2(T_c^2-T^2)/4$ as can be seen by equation (\ref{eq:Tc}).\\

 From (\ref{eq:Schrodinger}) and (\ref{eq:psi}), separating real and imaginary parts we have
 \begin{equation}
  \dot{\hat{\rho}}+\frac{\hbar}{m}(\hat{\rho}\Box^2 S+\boldsymbol{\nabla}S\boldsymbol{\nabla}\hat{\rho}
  -\dot{\hat{\rho}}\dot{S})=0\nonumber,
 \end{equation}
and
 \begin{eqnarray}
  -\frac{\hbar}{m}\dot S-\frac{\hbar^2}{2m^2}(\boldsymbol{\nabla} S)^2 &+&\frac{3\lambda}{2m^2}\hat{\rho}-\phi 
  +\lambda\frac{\hbar^2}{8m^2}k_B^2T^2\nonumber\\
  &+&\frac{\hbar^2}{2m^2}\left(\frac{\Box^2\sqrt{\hat{\rho}}}{\sqrt{\hat{\rho}}}\right)+\frac{\hbar^2}{2m^2}(\dot S)^2=0.\nonumber\\
 \label{hidro}
 \end{eqnarray}
Taking the gradient of (\ref{hidro}) and using the definition
 \begin{equation}
  \boldsymbol{v}\equiv\frac{\hbar}{m}\boldsymbol{\nabla}S
 \label{eq:vel}
 \end{equation}
then the velocity can be determined by the scalar function $S$, with this we obtain,
\begin{subequations}
 \begin{eqnarray}
  \dot{\hat{\rho}}+\boldsymbol{\nabla}\cdot(\hat{\rho}\boldsymbol{v})
  -\frac{\hbar}{m}(\hat{\rho}\dot{S}{\dot)}&=&0,\label{eq:cont}\\
  \dot{\boldsymbol{v}}+(\boldsymbol{v}\cdot\boldsymbol{\nabla})\boldsymbol{v}=-\boldsymbol{\nabla}\phi
  +\frac{3\lambda}{2m^2}\boldsymbol{\nabla}\hat{\rho}
  &+&\frac{\hbar^2}{2m^2}\boldsymbol{\nabla}\left(\frac{\nabla^2\sqrt{\hat{\rho}}}{\sqrt{\hat{\rho}}}\right)\nonumber\\
  +\frac{\hbar}{m}\dot{\boldsymbol{v}}\dot{\boldsymbol{S}}
  -\frac{\hbar^2}{2m^2}\boldsymbol{\nabla}\left(\frac{\partial^2_t\sqrt{\hat{\rho}}}{\sqrt{\hat{\rho}}}\right)
  &+&\frac{\lambda k^2_B}{4m^2}T\boldsymbol{\nabla}T\label{eq:2}
 \end{eqnarray}\label{eq:hydro}
\end{subequations}
Notice that in (\ref{eq:2}) $\hbar$ enters on the right-hand side through the term containing the gradient of $\hat{\rho}$. This term is usually called the 'quantum pressure' and is a direct consequence of the Heisenberg uncertainty principle, it reveals the importance of quantum effects in interacting gases.
Multiplying by $\hat{\rho}$, (\ref{eq:2}) can be written as:
 \begin{eqnarray}
  \hat\rho\dot{\boldsymbol{v}}+\hat\rho(\boldsymbol{v}\cdot\nabla)\boldsymbol{v}=\hat\rho\boldsymbol{F}_\phi -\boldsymbol{\nabla}p
  +\hat\rho\boldsymbol{F}_Q+\boldsymbol{\nabla}\cdot\sigma, 
 \label{eq:navier}
 \end{eqnarray}
where $\boldsymbol{F}_\phi=-\boldsymbol{\nabla}\phi$ is the force associated to the external potential $\phi$, $p$ can be seen as the pressure of the SF gas that satisfies the equation of state $p=w\hat\rho^2$, $\boldsymbol{\nabla}p$ are forces due to the gradients of pressure and $\omega=-3\lambda/4m^2$ is an interaction parameter, $\boldsymbol{F}_Q=-\boldsymbol\nabla U_Q$ is the quantum force associated to the quantum potential, \cite{b.q},
\begin{equation}
 U_Q=-\frac{\hbar^2}{2m^2}\left(\frac{\nabla^2\sqrt{\hat{\rho}}}{\sqrt{\hat{\rho}}}\right),
 \label{eq:UQ}
 \end{equation}
and $\boldsymbol{\nabla}\cdot\sigma$ is defined as
\begin{eqnarray}
 \boldsymbol{\nabla}\cdot\sigma=\frac{\hbar}{m}\hat{\rho}\dot{\boldsymbol{v}}\dot{S} 
 &+&\frac{1}{4}\frac{\lambda}{m^2}k_B^2\hat{\rho}T\boldsymbol{\nabla}T\nonumber\\
 &+&\zeta\boldsymbol{\nabla}(\ln\hat\rho{\dot)}
 -\frac{\hbar^2\hat\rho}{4m^2}\boldsymbol{\nabla}\left(\frac{\ddot{\hat\rho}}{\hat\rho}\right),\nonumber
 \label{eq:sigma}
 \end{eqnarray}
where the coefficient $\zeta$ is given by
\begin{equation}
 \zeta=\frac{\hbar^2}{4m^2}\left[-\boldsymbol\nabla\cdot(\hat{\rho}\boldsymbol{v})+\frac{\hbar}{m}(\hat{\rho}\dot{S})\dot{}\right]\nonumber,
 \label{eq:zeta}
 \end{equation}
and the term $\nabla(\ln\hat\rho{\dot)}$ can be written as
\begin{eqnarray}
 \boldsymbol{\nabla}(\ln\hat\rho{\dot)}=-\boldsymbol{\nabla}(\boldsymbol{\nabla}\cdot\boldsymbol{v})
 -\boldsymbol{\nabla}[\boldsymbol\nabla(\ln\hat{\rho})\cdot\boldsymbol{v}]
 +\frac{\hbar}{m}\boldsymbol{\nabla}[\frac{1}{\hat{\rho}}(\hat{\rho}\dot{S}{\dot)}]\nonumber
 \label{eq:vrho1}
 \end{eqnarray}
System \eqref{eq:hydro} is completely equivalent to equation (\ref{eq:Schrodinger}) and is the hydrodynamical representation of the latter.\\

 To simplify system (\ref{eq:hydro}) from now on we will neglect second order time derivatives and products of time derivatives. In 
this limit we arrive to the non-relativistic system of equations \eqref{eq:hydro}, which reads
\begin{subequations}
 \begin{eqnarray}
  \dot{\hat{\rho}}+\boldsymbol{\nabla}\cdot(\hat{\rho}\boldsymbol{v})&=&0,\label{eq:contNR}\\ 
  \hat\rho\dot{\boldsymbol{v}}+\hat\rho(\boldsymbol{v}\cdot\boldsymbol{\nabla})\boldsymbol{v}&=&\hat\rho\boldsymbol{F}_\phi 
  -\boldsymbol{\nabla}p+\hat\rho\boldsymbol{F}_Q+\boldsymbol{\nabla}\cdot\sigma.\nonumber\\
   \label{eq:navierNR}
 \end{eqnarray}\label{eq:hydroNR}
 \end{subequations}
Equation (\ref{eq:contNR}) is nothing but the continuity equation, and (\ref{eq:navierNR}) is the equation for the momentum it contains forces due to the external potential, to the gradient of the pressure, viscous forces due to the interactions of the condensate and finally forces due to the quantum nature of the equations.

 In this case $\boldsymbol{\nabla}(\ln\hat\rho{\dot)}$ now reads
\begin{equation}
 \boldsymbol{\nabla}(\ln\hat\rho{\dot)}=-\boldsymbol{\nabla}(\boldsymbol{\nabla}\cdot\boldsymbol{v})
 -\boldsymbol{\nabla}[\boldsymbol{\nabla}(\ln\hat{\rho})\cdot\boldsymbol{v}]\nonumber.
 \label{eq:vrho1NR}
 \end{equation}
 Thus
\begin{eqnarray}
 \boldsymbol{\nabla}\cdot\sigma=\frac{1}{4}\frac{\lambda}{m}k_B^2\hat{\rho}T\boldsymbol{\nabla}T\nonumber
 -\zeta\left[\boldsymbol{\nabla}(\boldsymbol{\nabla}\cdot\boldsymbol{v})
 +\boldsymbol{\nabla}[\boldsymbol{\nabla}(\ln\hat{\rho})\cdot\boldsymbol{v}]\right]\nonumber,
\label{eq:sigma2}
 \end{eqnarray}
where now we have
\begin{equation}
 \zeta=-\frac{\hbar^2}{4m^2}\boldsymbol{\nabla}\cdot(\hat{\rho}\boldsymbol{v})\nonumber,
 \end{equation}
Here we interpret the function $\boldsymbol{\nabla}\cdot\sigma$ as a viscosity expression, it contains terms which are gradients of the temperature and of the divergence of the velocity and density (dissipative contributions). The measurement of the temperature dependence in this thermodynamical quantity at the phase transitions might reveal important information about the behavior of the gas due to particle interaction.\\

In what follows we will derive the thermodynamical equations from the hydrodynamical representation. We can derive a conservation equation for a function $\alpha$, starting with the relationship
 \begin{equation}
  (\hat{\rho}\alpha)\dot{}=\hat{\rho}\dot{\alpha}+\alpha\dot{\hat\rho}
 \label{eq:phi1}
 \end{equation}
where $\alpha$ can take the values of $\phi$ and $U_Q$, both of them fulfil equation (\ref{eq:phi1}). Using the continuity equation (\ref{eq:contNR}) in \eqref{eq:phi1} we obtain,
  \begin{equation}
   (\hat{\rho}\alpha)\dot{}+\boldsymbol{\nabla}\cdot(\hat\rho\boldsymbol{v}\alpha)=-\hat\rho\boldsymbol{v}\cdot\boldsymbol{F}_{\alpha}
   +\hat\rho\dot{\alpha}\nonumber. 
   \label{eq:contphi}
  \end{equation}
Nevertheless, this procedure is not possible for $\sigma$ because in general we do not know it explicitly, only in some cases it might be possible to integrate it. 

 Observe how the quantum potential $U_Q$ also fulfills the following relationship
  \begin{equation}
   \hat\rho \dot U_Q+\boldsymbol{\nabla}\cdot(\hat{\rho}\boldsymbol{v}_\rho)=0,
   \label{eq:contUQ21}
  \end{equation}
which follows by direct calculation, and where we have defined the velocity density $\boldsymbol{v}_\rho$ by
  \begin{equation}
  \boldsymbol{v}_\rho=\frac{\hbar^2}{4m^2}\left(\boldsymbol{\nabla}\ln\hat\rho\right)\dot{}\nonumber,
  \label{eq:vrho}
  \end{equation}
which can be interpreted as a velocity flux due to the potential $U_Q$.  Using the continuity equation \eqref{eq:contNR}, equation (\ref{eq:contUQ21}) can be rewritten as
  \begin{equation}
   (\hat\rho U_Q)\dot{}+\nabla\cdot (\hat\rho U_Q\boldsymbol{v}+\boldsymbol{J}_\rho)+\hat{\rho}\boldsymbol{v}\cdot\boldsymbol{F}_Q=0
   \label{eq:contUQ2}
  \end{equation}
where we have  defined the quantum density flux 
  \begin{equation}
  \boldsymbol{J}_\rho=\hat{\rho}\boldsymbol{v}_\rho\nonumber.
   \label{eq:Jrho}
  \end{equation}
Equation \eqref{eq:contUQ2} is another expression for the continuity equation of the quantum potential $U_Q$.\\

 As we know, in general (for non-relativistic systems), the total energy density of the system $\epsilon$ is the sum of the kinetic, 
potential and internal energies \cite{b.r}, in this case we have an extra term $U_Q$ due to the quantum potential,
 \begin{equation}
  \epsilon=\hat{\rho}e=\frac{1}{2}\hat{\rho}v^2+\hat{\rho}\phi+\hat{\rho}u+\hat{\rho}U_Q
 \label{eq:energiae}
 \end{equation}
and where $u$ is the inner energy of the system. Then from (\ref{eq:energiae}) we have that $u$ will satisfy the equation
  \begin{equation}
   (\hat{\rho}u)\dot{}+\boldsymbol{\nabla}\cdot\boldsymbol{J}_u-\boldsymbol{\nabla}\cdot\boldsymbol{J}_{\rho}+\hat{\rho}\dot{\phi}=
   -p\boldsymbol{\nabla}\cdot\boldsymbol{v},
    \label{eq:conte}
  \end{equation}
being $\boldsymbol{J}_u$ the energy current,  given by a energy flux and a heat flux, $\boldsymbol{J}_q$,
  \begin{equation}
   \boldsymbol{J}_u=\hat\rho u\boldsymbol{v}+\boldsymbol{J}_q-p\boldsymbol{v},\nonumber
   \label{eq:Je}
  \end{equation} 
where $\boldsymbol{\nabla}\cdot\boldsymbol{J}_q=\boldsymbol{v}(\boldsymbol{\nabla}\cdot\sigma)$, expression that as we can see is related in a direct way to the velocity and gradients of temperature in the condensate, and is the one that shows in an explicit way the temperature dependence of the thermodynamical equations. With these definitions at hand we have,
  \begin{equation}
   \left(\hat\rho u\right)\dot{}+\boldsymbol{\nabla}\cdot(\hat\rho\boldsymbol{v}u+\boldsymbol{J}_q-p\boldsymbol{v}
   -\boldsymbol{J}_\rho)+\hat{\rho}\dot{\phi}=-p\boldsymbol{\nabla}\cdot\boldsymbol{v}.
   \label{eq:contu}
  \end{equation}

In order to find the thermodynamical quantities of the system in equilibrium (taking $p$ as constant on a volume $L$), we restrict the system to the regime where the auto-interacting potential is constant in time, with this conditions at hand for (\ref{eq:contu}) we have:
\begin{equation}
   \left(\hat\rho u\right)\dot{}+\boldsymbol{\nabla}\cdot(\hat\rho\boldsymbol{v}u+\boldsymbol{J}_q-p\boldsymbol{v}-\boldsymbol{J}_\rho)= 
   -p\boldsymbol{\nabla}\cdot\boldsymbol{v}
   \label{interna}
  \end{equation}
From (\ref{interna}) we can have a straightforward interpretation of the terms involved in the phase transition. As always the first term will represent the change in the internal energy of the system, $-p\boldsymbol{\nabla}\cdot\boldsymbol{v}$ is the work done by the pressure and $\boldsymbol{\nabla}\cdot\boldsymbol{v}$ is related to the change in the volume, $\boldsymbol{J}_q$ contains terms related to the heat generated by gradients of the temperature $\boldsymbol{\nabla}T$ and dissipative forces due to viscous forces $\sim\boldsymbol{\nabla}(\boldsymbol{\nabla}\cdot\boldsymbol{v})$ and finally but most important we have an extra term, $\boldsymbol{\nabla}\cdot\boldsymbol{J}_\rho$, due to gradients of the quantum potential (\ref{eq:UQ}).

 Integrating this resulting expression on a close region, we obtain
  \begin{eqnarray}
  \frac{\mathrm{d}}{\mathrm{d}t}\int\hat\rho u\,\mathrm{d}V+\oint (\boldsymbol{J}_q+p\boldsymbol{v})\cdot\boldsymbol{n}\, \mathrm{d}S
   &-&\oint\,\boldsymbol{J}_\rho\cdot\boldsymbol{n}\, \mathrm{d}S\nonumber\\ 
   &=&-p\frac{\mathrm{d}}{\mathrm{d}t}\int \,\mathrm{d}V\nonumber.
   \label{eq:contu2}
  \end{eqnarray}
Equation \eqref{interna} is the continuity equation for the internal energy of the system and as usual, from here we have an expression that would describe the thermodynamics of the system in an analogous way as does the first law of thermodynamics, in this case for the KG equation or a BEC. This reads
  \begin{equation}
   \mathrm{d}U=\text{\^d} Q+\text{\^d}A_Q-p\mathrm{d}V
   \label{eq:1leyBEC}
  \end{equation}
where as always, $U=\int \hat\rho u\,\mathrm{d}V$ is the internal energy of the system, \cite{b.m}, and as we can see, its change is due to a combination of heat $Q$ added to the system and work done on the system (pressure dependent), and
  \begin{equation}
   \frac{\text{\^d}A_Q}{\mathrm{d}t}=\frac{\hbar^2}{4m^2}\oint\,\hat\rho(\boldsymbol{\nabla}\ln\hat\rho)\dot{}\cdot\boldsymbol{n}\,
   \ \mathrm{d}S=\oint \hat\rho\boldsymbol{v}_\rho\cdot\boldsymbol{n}\,\mathrm{d}S,\nonumber
   \label{eq:AQ}
  \end{equation}
is the corresponding quantum heat flux due to the quantum nature of the KG equation. The second term on the right hand side of equation \eqref{eq:1leyBEC} would make the crucial difference between a classical and a quantum first law of thermodynamics.\\

From hereafter we study the transition between the $\Phi=0$ state to the low energy one with $T<T_c$ close to the minimum $\Phi_{min}=k_B\sqrt{T_c^2-T^2}/2$.\\

 During the time when $T>>T_c$ there are not scalar particles in the ground state. We will suppose that the 
scalar particles decouple from the rest of the matter at some moment, such that the total density here on remains constant. Below the critical temperature $T<T_c$, close to the local minimum the density oscillates around the value $\hat\rho=k_B^2 \kappa^2(T_c^2-T^2)/4$, being $ k_B \kappa T_c\sim 1$. The function $S$ in (\ref{eq:psi}) has a simple expression,  $S=s_0t$, with $s_0<<mc/\hbar$ in the non-relativistic case. This implies that the velocity $\boldsymbol{v}_0=0$, also, if there does not exist an external force in the system then, $\boldsymbol{F}_\phi=0$. In this case the viscosity (dissipative term) of the BEC might in fact contain the whole information of the phase transition.\\

 Finally, to illustrate the previous exposition, we give the following example. Suppose that in the system there are only condensed and 
excited particles of the same specie. Thus $\hat\rho=N_0/(N_{ex}+N_0)$ (total number of particles), where $N_0$ is the number of condensed particles and $N_{ex}$ the number of exited particles. Combining the equations obtained in the theoretical framework lead to the number of particles in the condensate in a four dimensional space,
  \begin{equation}
  N_0=\frac{N}{ k_B^2\kappa^2 T_c^2}\left[1-\left(\frac{T}{T_c}\right)^2\right],\nonumber
   \label{eq:N0}
  \end{equation}
being $N=N_0+N_{ex}$, and remembering we are considering space and time, note that in this case the exponent $2$ in the critical temperature appears naturally. As always this expression shows the dependence of the condensate fraction $N_0/N$ as a smooth function of temperature from $T\gtrsim T_c$ down to $T=0K$. In this case, the finite temperature terms are obtained from the one loop corrections of the SF density, and are down to be in complete agreement with the standard theory, \cite{b.s,b.t,b.u}.\\

 Observe that only a fraction $N/(k_B^2\kappa^2 T_c^2)$ of the scalar particles reaches the ground state at $T=0$, this value can only be 
determined experimentally and fits the value of the scale $\kappa$. So in principle we are able to mimic the result that in the presence of interactions we have $N_0<N$ even at $T=0$. The main idea we want to point out here is that these phenomena might be equivalent for a BEC on earth as for the cosmos, and this might follow the previous equations exactly, so this function might be tested in the laboratory. As the actually known observational evidence favors an open universe in what concerns BEC, we therefore need experimental tests to measure any new results that can be drown from the previous set of thermodynamical equations. If confirmed, the phase transition of a BEC can be explained using quantum field theory in a straightforward way.\\

 By adding a $U(1)$ SB term and temperature contributions to the effective Mexican hat potential of a system of weakly 
interacting bosons we have obtained several thermodynamical relations for the nonlinear Schr\"odinger equation (\ref{eq:Schrodinger}) beginning with KG. We have demonstrated another way, different from previous works, to obtain the equivalence of finite temperature SB and a BEC.

\acknowledgments
This work was partially supported by CONACyT M\'exico under grants 49865-F and I0101/131/07 C-234/07 of the Instituto Avanzado de Cosmologia (IAC) collaboration (http://www.iac.edu.mx/).


\begin{thebibliography}{99}

\bibitem{b.a}
  \Name{Pinto, M. B., Ramos, R. O. \and Parreira, J. E.}
  \REVIEW{Phys. Rev. D}{71}{2005}{123519}.
\bibitem{b.b}
  \Name{Griffin, A., Snoke, D. W. \and Stringari, S.}
  \Book{Bose-Einstein Condensation}
  \Editor{Pinto, M. B., Ramos, R. O. \and Parreira, J. E.}
  \Publ{Cambridge University Press}
  \Year{1995}.
\bibitem{b.c}
  \Name{Courteille, Ph. W., Bagnato, V. S. \and Yukalo, V. I.}
  \REVIEW{Laser Phys.}{11}{2001}{659}.
\bibitem{b.v}
  \Name{Haber, H. E. \and Weldon, H. A.}
  \REVIEW{Phys. Rev. Lett.}{46}{1981}{1497}.
\bibitem{b.d}
  \Name{Rivers, R. J.}
  \REVIEW{ArXiv}{}{2004}{cond-mat/0412404v1 preprint}.
\bibitem{b.e}
  \Name{Su\'arez, A. \and Matos, T.}
  \REVIEW{Mon. Not. R. Astron. Soc.}{416}{2011}{87}.
\bibitem{b.f}
  \Name{Chavanis, P. H. \and Harko, T.}
  \REVIEW{ArXiv}{}{2011}{astro-ph/1108.3986 preprint}.
\bibitem{b.g}
  \Name{Hung, C. -L., Zhang, X., Gemelke, N. \and Chin, C.}
  \REVIEW{Nature}{470}{2011}{236}.
\bibitem{b.h}
  \Name{Lin, Y. -J., Jim\'enez-Garc\'ia, K. \and Spielman, I. B.}
  \REVIEW{Nature}{471}{2011}{83}.
\bibitem{b.i}
  \Name{Plimak, L. I., Weib, C., Walser, R. \and Schleich, W. P.}
  \REVIEW{Optics Communications}{264}{2006}{311}.
\bibitem{b.j}
  \Name{Kirzhnits, D.A.}
  \REVIEW{JETP Lett.}{15}{1972}{745}.
\bibitem{b.k}
  \Name{Kolb, E. W. \and Turner, M. S.}
  \Book{The Early Universe}
  \Editor{Addison-Wesley}
  \Publ{William H. Press}
  \Year{1961}.
\bibitem{b.l}
  \Name{Dolan, L. \and Jackiw, R.}
  \REVIEW{Phys. Rev. D}{9}{1974}{3320}.
\bibitem{b.n}
  \Name{Pitaevskii, L. \and  Stringari, S.}
  \Book{Bose-Einstein Condensation}
  \Editor{Birman, J. \emph{et. al}}
  \Publ{Clarendon Press, Oxford}
  \Year{2003}.
\bibitem{b.o}
  \Name{Chiueh, T.}
  \REVIEW{Phys. Rev. E}{57}{2000}{4150}.
\bibitem{b.p}
  \Name{Bohm, D.}
  \REVIEW{Phys. Rev.}{85}{1952}{180}.
\bibitem{b.q}
  \Name{Gr$\ddot{o}$ssing, G.}
  \REVIEW{Phys. Lett. A}{388}{2009}{811}.
\bibitem{b.r}
  \Name{Oliver, X. \and de Sarac\'ibar, C. A.}
  \Book{Mec\'anica de Medios Continuos}
  \Editor{Vieira, E. \and Car E.}
  \Publ{UPC}
  \Year{2000}.
\bibitem{b.m}
  \Name{\"Ottinger, H. C.}
  \Book{Beyond Equilibrium Thermodynamics}
  \Editor{}
  \Publ{John Wiley \& Sons}
  \Year{2005}.
\bibitem{b.s}
  \Name{Haber, H. E. \and Weldon, H. A.}
  \REVIEW{Phys. Rev. D}{25}{1982}{502}.
\bibitem{b.t}
  \Name{Singh, S. \and Pathria, R. K.}
  \REVIEW{Phys. Rev. A}{30}{1984}{442}.
\bibitem{b.u}
  \Name{Singh, S. \and Pandita, P. N.}
  \REVIEW{Phys. Rev. A}{28}{1983}{1752}.







\end{thebibliography}
\end{document}